\documentclass[manuscript]{aastex}
\usepackage{natbib}
\usepackage{aas_macros}
\font\manual=manfnt at 7pt \def\dbend{\hbox{\raise0.9ex\hbox{\manual\char127\hspace{0.6em}}}}

\providecommand{\e}[1]{\ensuremath{\times 10^{#1}}}

\newcommand\Ion[2]{\ensuremath{\mathrm{#1\,\scriptstyle #2}}}

\newcounter{INTERNALionstage}
\providecommand{\ion}[2]{% replace the aastex version
  \setcounter{INTERNALionstage}{#2}%
  \Ion{#1}{\Roman{INTERNALionstage}}}

\def\gtsim{\mathrel{\hbox{\rlap{\hbox{\lower4pt\hbox{$\sim$}}}\hbox{$>$}}}}
\def\lesssim{\mathrel{\hbox{\rlap{\hbox{\lower4pt\hbox{$\sim$}}}\hbox{$<$}}}}

\def\g{{\rm\thinspace g}}

\def\K{{\rm\thinspace K}}

\def\htwo{\mbox{{\rm H}$_2$}}

\DeclareMathAlphabet{\vib}{OML}{cmm}{m}{it}
\begin{document}

%\title{Atomic data within the spectral synthesis code Cloudy}
\title{Stout: Cloudy's Atomic and Molecular Database}
\author{
M. L. Lykins\altaffilmark{1}, 
G. J. Ferland\altaffilmark{1,5}, 
R. Kisielius\altaffilmark{2}, 
M. Chatzikos\altaffilmark{1},
R. L. Porter\altaffilmark{6}, 
P. A. M. van Hoof\altaffilmark{3},
R. J. R. Williams\altaffilmark{4},
F. P. Keenan\altaffilmark{5},
P. C. Stancil\altaffilmark{6}
}

\altaffiltext{1}{University of Kentucky, Lexington, KY 40506, USA}

\altaffiltext{2}{Institute of Theoretical Physics and Astronomy, Vilnius 
                 University, A. Go{\v s}tauto 12, LT-01108, Lithuania}

\altaffiltext{3}{Royal Observatory of Belgium, Ringlaan 3, 1180 Brussels,
Belgium}

\altaffiltext{4}{AWE plc, Aldermaston, Reading RG7 4PR, UK}

\altaffiltext{5}{School of Mathematics and Physics, Queen's University Belfast, 
Belfast BT7 1NN, Northern Ireland, UK}

\altaffiltext{6}{Department of Physics and Astronomy and Center for Simulational 
Physics, University of Georgia, Athens, Georgia 30602-2451, USA}

\begin{abstract}
We describe a new atomic and molecular database we  developed for use in the 
spectral synthesis code Cloudy. The design of Stout is driven by the data needs 
of Cloudy, which simulates molecular, atomic, and ionized gas with kinetic 
temperatures $2.8 \K < T < 10^{10}~\K$ and densities spanning the low to 
high-density limits. The radiation field between photon energies
$10^{-8}$ Ry and 100 MeV is considered,
along with all atoms and ions of the lightest 30 elements, and $\sim 10^2$ molecules.
For ease of maintenance, the data are stored in a format as close as possible 
to the original data sources. Few data sources include the full range of data 
we need.  We describe how we fill in the gaps in the data or extrapolate rates 
beyond their tabulated range. We tabulate data sources both for the atomic 
spectroscopic parameters and for collision data for the next release of 
Cloudy. This is not intended as a review of the current status of atomic data,
but rather a description of the features of the database which we will build 
upon. 
\end{abstract}


\section{Introduction}

Cloudy is an openly available spectral simulation code based on detailed 
microphysics, most recently reviewed by \citet{CloudyReview13}. It considers 
microphysical processes from first principles to determine the excitation, 
ionization, and thermal properties of a mix of gas and dust. Much of this 
physics is described in \citet{AGN3}, hereafter AGN3. A very wide range of 
densities and temperatures can be modeled, and the full radiation between 
$10^{-8}$ Ry and 100 MeV is considered.

Massive amounts of atomic and molecular data are needed to do such simulations.
These include energy levels; transition probabilities; collision rates with 
electrons, protons, and atoms; photoionization cross sections; collisional 
ionization rate coefficients; recombination rate coefficients, along with 
charge exchange ionization/recombination data.
There are several spectral databases available, including Chianti 
\citep{Dere.K97CHIANTI---an-atomic-database-for-emission, Landi2012}
LAMDA \citep{Schoier.F05An-atomic-and-molecular-database-for-analysis}, 
JPL \citep{1998JQSRT..60..883P}, and CDMS \citep{Muller2001, Muller2005}. 
These provide energy levels, and transition probabilities. Chianti and LAMDA 
also include collision rates with a particular emphasis on certain applications.
Chianti and LAMDA are included in the Cloudy distribution and are used in our 
simulations
\citep{CloudyReview13,Lykins.M12Radiative-cooling-in-collisionally-and-photo}.

At times during the development of Cloudy, we have need to create additional 
models of atoms or molecules. What format should we use? Chianti comes closest 
to providing the data we need, but its format does not allow for more than 999 
levels and the collision rates are presented in a format that is far removed 
from the original published form. Only a few spline points are given for 
collision rates, and they emphasize temperatures higher than those found in 
photoionization equilibrium, so the fits are sometimes not valid at the low 
temperatures we need \citep{Ercolano2008}. Chianti's use of spline
interpolation can lead to unphysical
negative collision strengths in Chianti version 7. Furthermore, we must include 
collisions with atoms and molecules, which are important for 
photodissociation regions (PDR) calculations. Hence the 
need for our own database.

This paper describes how we implemented our spectral line database. It is not 
intended as a definitive reference for the state of the art in atomic and 
molecular data today. Continuous updates to the database will occur and be 
described in future papers.

\section{The Stout Database}

The new database was designed to have the following properties:
\begin{itemize}
\item The data format must be easy for a person to maintain since 
continual updating is necessary.
\item It must provide for different types of data. 
For example, radiative rates might be specified as oscillator  strengths, 
transition probabilities, or line strengths.
\item Collision data should, if possible, cover the temperature range considered 
by Cloudy,  currently 2.8 K to 10$^{10}$ K.  This is seldom available so we need 
to have a strategy to extrapolate beyond the limits of the tabulated data.
\item We must be able to reliably interpolate upon tables of rate coefficients
without producing unphysical negative values, which may introduce negative 
collision rates.
\item Both resonance and subordinate lines must be included since Cloudy is 
applied to dense environments where subordinate lines are important.
\item Both molecular and atomic data must be considered.
\item A broad range of collision partners, including electrons, \htwo, H$^0$, 
He, and H$^+$, must be considered.
\item Each file must explain its provenance by documentation at the end of the file.
\item As far as possible, the data must be presented 
in their original format.
We use the tabulated collision rates, collision strengths, energies, etc,
as they appear in the original publication.
This makes the data much easier to maintain.
\item Numbers within data files are free format.  Each number need only be
surrounded by a space or tab character to distinguish separate entries.
This makes it both easier to maintain and to remain close to the format of the 
original data source.
\item There must be no limit to the number of levels in a model.
\end{itemize}

\subsection{Spectral models in Cloudy}

We begin with a description of the atomic models in Cloudy. Cloudy has two 
distinctly different types of atomic models due to the different level 
structures of various isoelectronic sequences \citep{CloudyReview13}. 
The H- and He-like isoelectronic sequences have excited states that are closer 
to the continuum than the ground state. As a result, these states are strongly 
coupled to the continuum, with levels populated following recombination or 
collisions between excited states or the continuum. Excited states are 
relatively weakly coupling to the ground state.  The H- and He-like 
isoelectronic sequences use the models described by the series of papers 
starting with \citet{Bauman2005} and \citet{Porter2005}. 
\citet{Porter.R09Uncertainties-in-theoretical-HeI-emissivities:-HII-regions,
Porter.R12Improved-He-I-emissivities-in-the-case-B-approximation,
2013MNRAS.433L..89P} give the most recent updates.

For more complex ions, the lowest excited states are close to the ground state 
and are strongly coupled to it. The influence of the continuum is weak.
The remaining atoms and ions are treated with the atomic models described in 
this paper or with Chianti. The Appendix contains Table~\ref{TabDataSources} 
that summarizes the data sources we use. We describe how we use these data in 
the following subsections.

The data for the atomic species, contained in the Stout database, are located in
separate directories for each species with a structure similar to that of Chianti. 
Each data set consists of three files - energy levels (the file with the 
extension ``nrg''), transition probabilities (extension ``tp''), and collision 
data (extension ``coll'').

\subsection{Energy levels}
\label{lev}

We use the experimental level energies from NIST \citep{NIST_ASD} if possible. 
Experimental data are available for most species.  The level energies given 
in the NIST database are usually derived from  measured line wavelengths. 
If there are no experimental data, we utilize theoretical data which often are 
of lower accuracy.

Transition rates that come from theoretical calculations must be corrected for 
any differences between experimental and theoretical energies. The level 
ordering may not agree so it is absolutely important to match the level 
assignments given in different data sources. When this issue is overcome, the 
integrity of the particular system is assured, and the calculated collisional 
parameters and radiative parameters are consistent with the experimental energy 
levels. 

By default we report wavelengths, in {\AA}ngstr{\"o}m that are derived from 
the stored energy levels (so-called Ritz wavelengths). We do allow the 
wavelength to be specified to override this default. The convention in atomic 
physics is to use air wavelengths for $\lambda > 2000$~\AA\ and vacuum 
wavelengths for $\lambda < 2000$~\AA. More recently, work has started to appear 
which uses vacuum for all wavelengths. The Sloan project (see The Sloan Digital Sky Survey at 
http://www.sdss.org) is an example. By default we follow the atomic physics 
convention but provide an option to report only vacuum wavelengths. The index 
of refraction of air is taken from \citet{Peck.E72Dispersion-of-Air}.

%%%%%%%%%%%%%% Table level %%%%%%%%%%%%%%%%%%%%%
\renewcommand{\baselinestretch}{1.0}
\begin{table}[!t]%[ht]
\caption{
\label{levels}
A sample of the level energies file (s\_2.nrg) for  \ion{S}{2}
from Stout. The first column represents the level index that is used in 
the transition probability and collision data files, the second column
gives level energies (in cm$^{-1}$), the third column gives level
statistical weights $g$, and the last two columns give a level
designation with ``*'' standing for odd-parity levels.
}
\flushleft
%\large
\begin{tabular}{lllll}
\hline
N& energy &$g$ & configuration& $LS\pi$ \\
\hline
1 &	0.00	    &	 4	 &	3s2.3p3	 &	4S*  \\	
2 &	14852.94 &		4	 &	3s2.3p3	 &	2D*  \\	
3 &	14884.73 &		6	 &	3s2.3p3	 &	2D*  \\	
4 &	24524.83 &		2	 &	3s2.3p3	 &	2P*  \\	
5 &	24571.54 &		4	 &	3s2.3p3	 &	2P*  \\	
6 &	79395.39 &		6	 &	3s.3p4	  &	4P   \\	
7 &	79756.83 &		4	 &	3s.3p4	  &	4P   \\	
\hline
\end{tabular}
\end{table}
%%%%%%%%%%%%%%%%%%%%%%%%%%%%%%%%%%%%%%%%%%%%%%%%%%%%%%%%

We present a sample of a file with columns description for the energy level 
data in Stout in Table~\ref{levels}. Just a small part of the file s\_2.nrg 
with the \ion{S}{2} ion level energies are given here. The complete data table 
is given by \citet{Kisielius2014}, whereas level data are taken from the NIST 
database \citep{NIST_ASD}. Only the level numbers, energies, and the statistical
weights are utilized for deriving of the transition probability or collision 
strengths, whereas the configuration and $LS\pi$ are given for information
purposes.

\subsection{Radiative transitions}
\label{tran}

The radiative transition between the upper level $u$ and the lower level $l$ 
can be parameterized as a line strength $S$, oscillator strength $f$ (or $gf$), 
or transition probability $A$, although only the latter enters in a calculation 
of level populations and emission spectra. Different data sources will provide 
different parameters, and we accept all three.

We prefer to utilize the transition line strength $S$ over the weighted 
oscillator strengths $gf$ or transition probabilities $A$. The advantage to $S$ 
is that it does not depend explicitly on the transition energy $\Delta E$ (or 
the transition wavelength $\lambda$), whereas $gf$ and $A$ do. Many published 
transition data are the result of theoretical calculations and use theoretical 
energies while we use experimental energies where possible. Therefore, a 
correction due to the uncertainty in the calculated transition energy 
$\Delta E$, or wavelength $\lambda$, values must be done. The conversion to 
the experimental transition energies $\Delta E_{\mathrm{exp}}$ or observed 
transition wavelengths $\lambda_{\mathrm{exp}}$ is: 
\begin{equation}
A_{\mathrm{corr}}^k = 
A_{\mathrm{th}}^k (\Delta E_{\mathrm{exp}}/\Delta E_{\mathrm{th}})^{2k+1}
=A_{\mathrm{th}}^k (\lambda_{\mathrm{th}}/\lambda_{\mathrm{exp}})^{2k+1}
\label{eq:corr}
\end{equation}
where $k$ is the transition multipole order ($k=1,2,3,...$), $A$ is the
transition probability, $\lambda_{\mathrm{th}}$ is the theoretical transition
wavelength. 

%%%%%%%%%%%%%%%%%%%%%%%%%%%%%%%%%%%%%%%%%%%%%%%%%%%%%%%%%%%%%%%%%%%%%%%%%%%%%%%%%
\renewcommand{\baselinestretch}{1.5}
\begin{table}[!t]%[ht]
\caption{Conversion factors and coefficients from the transition line strengths $S$ 
(in a.u.) to the radiative transition probabilities $A$ (in s$^{-1}$).
\label{conversion}
}
\centering
\begin{tabular}{llcc}
\hline
type& factor $C_{\lambda}$ & $C_{\lambda}$& $C_{\Delta E}$ \\
\hline
$E1$ & $\frac{64 \pi^4 e^2 a_0^2}{3h}$                    &$2.02613\e{18}$ & $2.14200\e{10}$\\
$M1$ & $\frac{64 \pi^4 \mu_{\mathrm B}^2}{3h}$            &$2.69735\e{13}$ & $2.85161\e{5} $\\
$E2$ & $\frac{64 \pi^6 e^2 a_0^4}{15h}$                   &$1.11995\e{18}$ & $5.70322\e{4} $\\
$M2$ & $\frac{64 \pi^6 \mu_{\mathrm B}^2 a_0^2}{15h}$     &$1.49097\e{13}$ & $7.59260\e{-1}$\\
$E3$ & $\frac{2048 \pi^8 e^2 a_0^6}{4725h}$               &$3.14441\e{17}$ & $7.71311\e{-2}$\\
$M3$ & $\frac{2048 \pi^8 \mu_{\mathrm B}^2 a_0^4}{4725h}$ &$4.18610\e{12}$ & $1.02683\e{-6}$\\
\hline
\end{tabular}
\end{table}
%%%%%%%%%%%%%%%%%%%%%%%%%%%%%%%%%%%%%%%%%%%%%%%%%%%%%%%%%%%%%%%%%%%%%%%%%%%%%%%%%

The transition line strength $S$ is expressed in atomic units (a.u.). It is 
symmetric in relation to the initial and final states, and is obtained as a 
square of the corresponding $E1$, $M1$, $E2$, $M2$, $E3$, $M3$ transition matrix 
elements. In this case, the electric multipole emission transition probability 
(Einstein $A$-coefficient) 
$A_{ul}^k$ (in s$^{-1}$) can be determined as
\begin{equation}
A_{ul}^k = C_{\lambda} S /( g_u \lambda^{2k+1})
\label{eq:cl}
\end{equation}
where $g_u$ is the statistical weight of the upper level, and $C_{\lambda}$ is 
the conversion factor. The expressions of the factor $C_{\lambda}$ for various 
multipole transitions are presented in Table~\ref{conversion}. We provide 
numerical values of the conversion coefficients when $\lambda$ is expressed 
in \AA\ and the line strength $S$ is calculated in a.u. In the case when we 
have the transition energy $\Delta E$ instead of wavelength, we can use similar 
expression for 
$A_{ul}^k$:
\begin{equation}
A_{ul}^k = C_{\Delta E} (\Delta E)^{2k+1} S / g_u .
\label{eq:ce}
\end{equation}
The conversion coefficients $C_{\Delta E}$ are given in Table~\ref{conversion}
for $\Delta E$ determined in a.u. For different energy units, one must
rely on these standard relations: 
1 a.u.$=2$ Ry$=27.211385$ eV$=219474.63$ cm$^{-1}$, and the inverse relations:
1 Ry$=0.5$ a.u.; 1 eV$=0.036749324$ a.u.; 1 cm$^{-1}=4.5563353 \cdot 10^{-6}$ 
a.u.
Having the radiative transition probabilities of various multipole orders, one 
can simply derive the absorption oscillator strengths $f_{lu}$ using a simple 
expression:
\begin{equation}
f_{lu}= 1.4992\e{-16} \lambda^2 (g_u/g_l) A_{ul}
\label{osc}
\end{equation}
where $g_{l}$ is the statistical weight of the lower level. The relation between 
the oscillator strength $f$ and transition probability $A$ does not depend on 
the radiative transition type.

%%%%%%%%%%%%%% Table transitions %%%%%%%%%%%%%%%%%%%%%
\renewcommand{\baselinestretch}{1.0}
\begin{table}[!t]%[ht]
\caption{
\label{tp}
A sample of the transition probabilities file (s\_2.tp) for \ion{S}{2}
from Stout. The first column represents radiative transition data type
(``A'' for a transition probability, ``f'' for weighted oscillator strength, 
``S'' for a line strength), the second column gives the lower level index, 
the third column gives the upper level index, the fourth column gives  
a transition parameter value, and the final column points to the radiative
transition type ($E1$, $E2$, $E3$, ..., $M1$, $M2$, ...).
}
\flushleft
%\large
\begin{tabular}{lllll}
\hline
data &  & & & transition \\
type& $N_l$ &$N_u$ & TP & type \\
\hline
S &		1	&	2	&	5.54E-03	&	E2 \\
S &		1	&	2	&	1.77E-05	&	M1 \\
S &		1	&	3	&	1.29E-02	&	E2 \\
S &		1	&	3	&	6.30E-07	&	M1 \\
S &		1	&	4	&	2.25E-06	&	E2 \\
S &		1	&	4	&	3.37E-04	&	M1 \\
S &		1	&	5	&	3.00E-10	&	E2 \\
S	&  1	& 5	& 1.67E-03	& M1 \\
S	&  1	& 6	& 2.65E-01	& E1 \\
\hline
\end{tabular}
\end{table}
%%%%%%%%%%%%%%%%%%%%%%%%%%%%%%%%%%%%%%%%%%%%%%%%%%%%%%%%

Table~\ref{tp} gives a sample of the transition data file for  \ion{S}{2}, with 
the data coming from \citet{Kisielius2014}. The transition line strengths $S$ 
are given as the basic radiative transition data as they do not depend 
explicitly on the transition energy. The NIST database traditionally provides 
the radiative transition probabilities (rates) $A$. Conversion from the 
transition line strengths $S$ to the transition probabilities $A$ depends on 
the transition type. It can be performed with the help of Table~\ref{conversion}.

\subsection{Collisions}
%\subsection{Collision rates}
\label{coll}

Collisional data can be given as collision strengths, effective collision 
strengths, collision cross sections, and rate coefficients. In the 
electron-impact excitation:
\begin{equation}
\mathrm{A}^{N+}(E_l) + e(\varepsilon_l) \rightarrow 
\mathrm{A}^{N+}(E_u) + e(\varepsilon_u),
\label{excit}
\end{equation}
the energy conservation law leads to
\begin{equation}
E_l+\varepsilon_l = E_u+\varepsilon_u .
\label{energy}
\end{equation}
Here $E_l$ and $E_u$ are the energies of the lower and upper levels, 
$\varepsilon_l$ and $\varepsilon_u$ are the kinetic energies of the incident 
and the scattered electron.  

Maxwellian-averaged effective collision strengths are utilized in Stout. Here 
we provide the basic relations for these parameters, whereas data sources are 
given in Table~\ref{TabDataSources}. Our preferred method is to use collision 
data calculated in the close-coupling (CC) approach, e.g., the R-matrix method. 
Unfortunately, such data are not available for many ions. Even in the case 
when some data exist, these usually deal with only a few of the lowest levels 
or even the $LS$ terms with unresolved fine-structure levels. So one must 
resort to less elaborate approaches, such as the distorted-wave method, the 
plane-wave approximation, or a $\bar g$ (g-bar) formula, for the remaining data.

The dimensionless collision strength $\Omega$ is the best to describe the 
electron-impact excitation process from the lower level $E_l$ to the excited 
level $E_u$. It is symmetrical in regard to the initial and final states 
parameter, i.e, $\Omega_{lu} = \Omega_{ul}$. For ions, it has a finite value at 
the excitation threshold and varies only slightly with the incident electron 
energy if autoionization resonances are not considered. For neutral atoms, 
the collision strength $\Omega_{lu}$ goes to zero at the excitation threshold. 

At high incident electron energies, the behavior of the collision strength 
depends on the transition (line) type and has a different form for allowed or 
forbidden transitions. The collision strength $\Omega_{lu}$ is determined as a 
square of the excitation operator's matrix element. It is connected to the 
excitation cross-section $\sigma_{lu}$ and the de-excitation cross-section 
$\sigma_{ul}$ by simple relations:
\begin{equation}
\sigma_{lu}(\varepsilon_l) = \Omega_{lu} \frac{\pi \mathrm{a}_0^2}{g_l E_l}
\label{sigma_lu}
\end{equation}
and
\begin{equation}
\sigma_{ul}(\varepsilon_u) = \Omega_{ul} \frac{\pi \mathrm{a}_0^2}{g_u E_u},
\label{sigma_ul}
\end{equation}
where $\pi \mathrm{a}_0^2 = 8.7972\e{-17}$ cm$^2$\, is the atomic cross-section 
unit. For  electric dipole allowed transitions, one can express the excitation
cross section $\sigma_{lu}$ through an effective Gaunt factor 
${\bar g}_{lu}(\varepsilon_l)$ (as in \citealp{Mewe1972}),
often called the ``g-bar approximation'':
\begin{equation}
\sigma_{lu}(\varepsilon_l)=\frac{8\pi f_{lu}{\bar g}_{lu}(\varepsilon_l)}
{\sqrt{3} \varepsilon_l (E_u-E_l)}
\label{gbar}
\end{equation}
with $f_{lu}$ being the absorption oscillator strength.

The collision strengths are integrated over a Maxwellian distribution of 
free-electron energies in order to determine the effective collision strengths
$\Upsilon_{lu}$ = $\Upsilon_{ul}$ (or rate parameters) at some electron 
temperature $T_e$:
\begin{equation}
\Upsilon_{lu}=\int_0^{\infty} \Omega_{lu}(\varepsilon_u) 
\exp(-\varepsilon_u/k_{\mathrm{B}}T_e) d(-\varepsilon_u/k_{\mathrm{B}}T_e).
\label{upsilon}
\end{equation}
Here $k_{\mathrm{B}}$ refers to Boltzmann's constant. In this case, the 
excitation rate coefficient $q_{lu}$ (in cm$^3$~s$^{-1}$) is expressed as
\begin{equation}
q_{lu}=8.629\e{-6}\frac{ \exp((E_l-E_u)/k_{\mathrm{B}}T_e) \Upsilon_{lu}}{\g_l T_e^{1/2}},
\label{rate}
\end{equation}
whereas the de-excitation rate coefficient $q_{lu}$ is determined by formula:
\begin{equation}
q_{ul}=8.629\e{-6}\frac{\Upsilon_{lu}}{\g_u T_e^{1/2} }.
\label{rk_rate2}
\end{equation}

%%%%%%%%%%%%%% Table collisions %%%%%%%%%%%%%%%%%%%%%
\renewcommand{\baselinestretch}{1.0}
\begin{table}[!t]%[ht]
\caption{
\label{t_coll}
A fragment of the effective collision strength file \texttt{s\_2.coll} for the 
ion \ion{S}{2} from Stout. The first column represents data type, e.g.,
``TEMP'' stands for a temperature grid (in \K), ``CSELECTRON'' for the
effective electron-impact excitation strength $\Upsilon$, ``RATE PROTON" 
for the proton excitation rate, etc. The data types and their sources are
provided in the same file in the comments lines. The second column gives the 
lower level index, the third column gives the upper level index, and next
columns give a particular collision parameter for the corresponding temperature.
}
\flushleft
%\large
\begin{tabular}{llllllll}
\hline
data &  & & &&&&\\
type& $N_l$ &$N_u$ &   &&&&\\
\hline
TEMP       &	  &   & 5.00E$+$03 &	7.00E$+$03 &	1.00E$+$04 &	1.50E$+$04 &	2.00E$+$04\\	
CSELECTRON &	1 &	2 &	2.66E$+$00 &	2.62E$+$00 &	2.56E$+$00 &	2.48E$+$00 &	2.41E$+$00\\	
CSELECTRON &	1 &	3 &	3.98E$+$00 &	3.91E$+$00 &	3.83E$+$00 &	3.71E$+$00 &	3.61E$+$00\\	
CSELECTRON &	1 &	4 &	6.86E$-$01 &	6.94E$-$01 &	7.04E$-$01 &	7.17E$-$01 &	7.27E$-$01\\	
CSELECTRON &	1 &	5 &	1.38E$+$00 &	1.39E$+$00 &	1.42E$+$00 &	1.44E$+$00 &	1.46E$+$00\\	
CSELECTRON &	1 &	6 &	2.25E$+$00 &	2.36E$+$00 &	2.54E$+$00 &	2.75E$+$00 &	2.84E$+$00\\	
CSELECTRON &	1 &	7 &	2.01E$+$00 &	2.09E$+$00 &	2.19E$+$00 &	2.30E$+$00 &	2.31E$+$00\\	
\hline
\end{tabular}
\tablecomments {
Either deexcitation rate coefficients (cm$^3$ s$^{-1}$) 
or effective collision strengths can be specified.
The colliders include electrons, protons, alpha particles, He$^+$, He$^0$, 
\htwo\ (ortho and para), and H$^0$.
}
\end{table}
%%%%%%%%%%%%%%%%%%%%%%%%%%%%%%%%%%%%%%%%%%%%%%%%%%%%%%%%

Table~\ref{t_coll} gives a fragment of the collision data file s\_2.coll for 
\ion{S}{2}. For this ion we employ the electron-impact excitation data from
\cite{Tayal2010}.
There can be several temperature grids in one data file, especially when
different projectiles, such as electrons, protons, hydrogen atoms, or hydrogen
molecules, are described. In a similar way, there can be different data sources
for different level combinations even for the same collider.
The collision data for a given transition and collider will be overwritten
if new data for that transition and collider appear
later in the file when read in by Cloudy.

Most of the data sources in Table~\ref{TabDataSources} are given for the 
electron collisions. Nevertheless, some colliders other than electrons
are included. 
For an atomic hydrogen collider, very important in PDRs, we use data from 
\citet{Launay1977} for  \ion{C}{1}, from \citet{Abrahamsson2007, Krems2006}
for \ion{O}{1}, from \citet{Hollenbach1989} for  \ion{Ne}{2}, from
\citet{Barklem2012} for \ion{Mg}{1}, and from \citet{Barinovs2005} for 
\ion{Si}{2}. For  proton colliders, very important in collisions between
levels with similar energies, we utilize data from 
\citet{Roueff1990} for \ion{C}{1} and data from \citet{Pequignot1990} for
\ion{O}{1}. For H$_2$, data from \citet{Schroder1991} are 
employed for \ion{C}{1}, data from \citet{Wiesenfeld2014} for the \ion{C}{2}
ion, and data from \citet{Jaquet1992} for \ion{O}{1}. For  helium impact
on neutrals, we use data from \citet{Staemmler1991} for \ion{C}{1}, and from
\citet{Monteiro1987} for \ion{O}{1}.

For collisions involving molecules, the literature often gives deexcitation 
rates rather than collision strengths. We accept deexcitation rates for any 
transition and species.

It is necessary to interpolate within tables of collision rates versus 
temperature, and in many calculations, extrapolate beyond the tabulated range.
Within the table, we interpolate using the method of \cite{Fritsch84},
which is local and piecewise cubic, and maintains the monotonicity properties 
of the underlying data. This ensures that the interpolation does not introduce 
any ``overshoots'', where the interpolated value does not lie within the range 
of the tabulated data. Such overshoots appear to be the source of the negative 
collision strengths that are present in version 7 of the Chianti database.

\subsubsection{Temperature extrapolation for atoms and ions}

Cloudy considers the temperature range extending from 2.8 K to 10$^{10}$~K 
and considers all ions of the first thirty elements along with several dozen 
molecules.  Gaps in the collision data are common. Often we must extrapolate 
beyond the range of the tabulated data, or improvise entire collections of data.

For temperatures below the range of the tabulated data, and for ions with 
positive charge, we assume that the collision strength is constant to 
extrapolate below the lowest tabulated temperature.  Physically, an effective 
collision strength is a Boltzmann average over the excitation cross section. 
As the temperature goes to zero, this average is over a narrow range near 
threshold, and will tend to be constant. This is not true if there are strong 
resonances very near threshold but it is a reasonable first approximation.  
For neutral species the collision strength goes to zero at energies near
threshold so the effective collision strength also goes to zero as 
$T \rightarrow 0$. We do a linear interpolation between the lowest temperature 
value and 0.0. We use these collision strength laws to form the appropriate 
temperature scaling when working with data giving collisional deexcitation 
rates.

For high temperatures we use \citet{BurgessTully1992} to guide the extrapolation.
\citet{BurgessTully1992} consider three possible types of transitions with 
different behavior at high energies (temperatures), Type 1 for the electric 
dipole transitions, Type 2 for the non-electric dipole, non-exchange 
transitions, Type 3 for the exchange transitions, with Types 4 and 5 being 
special cases (for more details on the transition classification see 
\cite{BurgessTully1992}). We use the first two types, Type 1 and Type 2. 
The Type 3 classification can be avoided when levels and transitions
are expressed in the intermediate coupling rather than pure $LS$ coupling.
In this case one can not separate the spin-changing transitions since the 
selection rules are applied for the total angular momentum $J$.

Transition types can be deduced from the energy levels files (*.nrg) where
$J$, configurations and their parities are given or from the radiative
transition files (*.tp) where transition types are given (but these are
not present in all transition data files). In general, our data files contain
information necessary to make a separation between Type 1 and Type 2 transitions.
For the Type 1 transitions, the high-temperature behavior of the effective
collision strength $\Upsilon$ is described by a simple relation 
$\Upsilon = C_1 \ln(T_e)$. The value of $C_1$ can be derived from the last
tabulated temperature point in the data file. For the Type 2 transitions, 
the effective collision strength does not depend on the electron temperature
$T_e$, i.e., $\Upsilon =C_2$. The value of $C_2$ is the value of $\Upsilon$ at
the last tabulated temperature.

Tests show that the low-temperature extrapolation does affect calculations.
In photoionization equilibrium very low kinetic temperatures are possible
\citep{AGN3}. The constant temperature test cases 
in the Leiden PDR comparison \citep{Roellig2007} have $T_{\mathrm{kin}} = 50\K$, 
lower than many tabulated rates. Predictions of some Leiden test cases were 
affected by the form of the low-$T$ extrapolation.

\subsubsection{Gaps in the collision data}
\label{section:GAPS}

We use the $\bar g$ approximation \citep{Seaton1962,VanRegemorter1962} to fill 
in missing electron collision data. This is a highly approximate relationship 
between the transition probability or oscillator strength, and the collision 
strength. We use \citet{Mewe1972} for those isoelectronic sequences he 
considered, and \citet{VanRegemorter1962} for others.

We provide a way to test the effects of such uncertain data.
Cloudy includes a built-in Gaussian random noise generator.
This was used, for instance, to assay the effects of missing
\htwo\ collision rates upon the final spectrum 
\citep{Shaw.G05Molecular-Hydrogen-in-Star-forming-Regions}.
Repeated calculations will reveal the uncertainties introduced by
the approximations, if the uncertainties can be quantified.

Some databases have no radiative transition between large blocks of levels.
For instance, a species may have no E1 transitions between the ground
and first excited configurations.  Higher order transitions are possible but 
many databases present only E1 transitions.  If theoretical collisional rates
have not been computed, then there would not be any coupling between the configurations.
It is not possible to simultaneously solve for the populations; the matrix becomes
ill conditioned.
In cases where we have no radiative or collision data, we leave the radiative 
transition rate as zero and use an electron effective collision strength of $10^{-10}$ .
This was chosen to be as small as possible while allowing the
linear algebra to function properly.

\subsubsection{Temperature extrapolation for molecular excitation}

When molecular collisional deexcitation rate coefficients $q(T)$ are provided 
only over a limited temperature range, the following two simple extrapolation 
approaches are applied:  
\begin{equation}
q(T) = q(T_{\rm low}), ~~~~~~T < T_{\rm low},
\label{lowT}
\end{equation}
and
\begin{equation}
q(T) = q(T_{\rm high})\left ({{T}\over{T_{\rm high}}} \right )^{1/2} \exp(-T/10^5 {~\rm K}), ~~~~~T > T_{\rm high}.
\label{highT}
\end{equation}
$T_{\rm low}$ and $T_{\rm high}$ correspond to the low and high temperature 
limit, respectively, of the data. The extrapolation formulae are valid for 
inelastic collisions of neutral molecules (e.g., OH) or molecular ions 
(e.g., HCO$^+$) with neutral colliders (e.g., H, He, H$_2$) for deexcitation 
(downward) transitions resulting in changes in fine-structure, rotational, 
and/or vibrational levels of the target molecules. Physical justifications for 
the extrapolations as well as caveats for their use are described below.

\subsubsubsection{Extrapolation to low temperature}  

The deexcitation rate coefficients as a function of temperature $T$ are 
obtained by thermally averaging the inelastic integral cross sections over a 
Maxwellian kinetic energy distribution given by
\begin{equation} \label{rate1}
q_{u\rightarrow l}(T) = \left (\frac{8k_{\mathrm{B}}T}{\pi \mu} \right )^{1/2}
\frac{1}{(k_{\mathrm{B}}T)^{2}}\int_{0}^{\infty}\sigma_{u\rightarrow l}(E_{\mathrm{kin}})
\exp(-E_{\mathrm{kin}}/k_{\mathrm{B}}T)E_{\mathrm{kin}}dE_{\mathrm{kin}},
\end{equation}
where $\sigma_{u\rightarrow l}(E_{\mathrm{kin}})$ is the state-to-state inelastic cross 
section, $E_{\mathrm{kin}}$ the center of mass kinetic energy, $\mu$ the reduced mass 
of the collision complex,  and $l$ ($u$) the lower (upper) levels in
the molecule.

Rewriting Eq.~(\ref{rate1}) with the cross section in terms of the relative 
velocity $v$ of the collision system gives
\citep{flo90}
\begin{equation} \label{rate2}
q_{u\rightarrow l}(T) =  \left (\frac{2}{\pi} \right)^{1/2} 
\left (\frac{\mu}{k_{\mathrm{B}}T} \right)^{3/2}
\int_{0}^{\infty}\sigma_{u\rightarrow l}(v)
\exp(-\mu v^{2}/2k_{\mathrm{B}}T)v^{3}dv.
\end{equation}
If the cross section is assumed to have the analytical form
\begin{equation} \label{sigma}
\sigma_{u\rightarrow l}(v) = Bv^{a},
\end{equation}
for all $v$ (or $E_{\mathrm{kin}}$)
where $B$ is an (undetermined) constant and $a$ is some power, then the rate 
coefficient takes the form
\begin{equation} \label{krate}
q(T)=A(a)B(T/\mu)^{b}
\end{equation}
\citep{wal14}. Here $b=(a+1)/2$ and $A$ is a function of $a$, both deduced 
from the Gaussian integral in Eq.~(\ref{rate2}). This result is exact, 
given the assumption of Eq.~(\ref{sigma}), and applicable to all collision 
systems. It is approximate if the cross section dependence varies with $v$ as 
in real systems.

Now at sufficiently low kinetic energy, \citet{wig48} showed that the inelastic 
cross section takes the form
\begin{equation}
\sigma_{u\rightarrow l} \sim E^{\ell-1/2}_{\mathrm{kin}},
\end{equation}
where $\ell$ is the total orbital angular momentum of the collision complex. 
In most systems of interest, $s$-wave scattering (i.e., $\ell=0$) is allowed and 
dominates at low kinetic energy. Therefore,
\begin{equation}
\sigma_{u\rightarrow l} \sim E^{-1/2}_{\mathrm{kin}} \sim v^{-1},
\label{Wigner}
\end{equation}
or $a=-1$, $b=0$ and the rate coefficient becomes a constant, independent of 
temperature as given by Eq.~(\ref{lowT}). 

Equation~(\ref{lowT}) is absolutely valid under two conditions: i) when 
$T < T_{\rm Wigner}$, the so-called Wigner regime, where higher partial waves 
($\ell >0$) do not contribute to the cross section and ii) when relevant 
selection rules do not forbid $s$-wave scattering. Typically, 
$T_{\rm Wigner} \sim$ mK and much less than $T_{\rm low}$. However, for 
$T\sim 1-100$~K the rate coefficient is usually  oscillatory due to the presence 
of orbiting and Feshbach resonances in the cross section. This behavior cannot 
be easily analytically reproduced so that extrapolating the Wigner threshold 
behavior of Eq.~(\ref{Wigner}) to $T_{\rm low}$ is a reasonable 
pragmatic approach.

In the event that $s$-wave scattering is forbidden for the particular 
transition (which is rare), the cross section would drop rapidly to zero as $v$ 
goes to zero
\begin{equation}
\sigma_{u\rightarrow l} \sim E^{1/2}_{\mathrm{kin}} \sim v.
\label{p-wave}
\end{equation}
The rate coefficient would be overestimated by Eq.~(\ref{lowT}), but this 
error would be limited to low astrophysical temperatures,
$T \lesssim$ 10 K.

\subsubsubsection{Extrapolation to High Temperature} 

A number of approaches have been proposed for extrapolating the deexcitation 
rate coefficient to higher temperatures beyond $T_{\rm high}$. For linear 
molecules, \citet{Schoier.F05An-atomic-and-molecular-database-for-analysis} 
fitted the available data in LAMDA to the form
\begin{equation}
q(T) = {{\alpha}\over {k_{\mathrm{B}}}T}\exp(-\beta/(k_{\mathrm{B}}T)^{1/4})
\exp(-\gamma/(k_{\mathrm{B}}T)^{1/2})
\label{fit}
\end{equation}
where $\alpha$, $\beta$, and $\gamma$ are fit parameters and then used the fit 
for $T>T_{\rm high}$. A more pragmatic approach, which avoids fitting, is to 
apply a hard-sphere model. This assumes that the cross section is independent of 
kinetic energy giving $a=0$ and $b=1$ in Eqs.~(\ref{sigma}) and 
(\ref{krate}), respectively, so that
\begin{equation}
q(T) \sim T^{1/2}.
\label{highrate}
\end{equation}
However, the inelastic cross section typically turns up to a maximum near a few 
eV before decaying at higher energies due to the increasing importance of 
collisional dissociation, electronic excitation, and collisional ionization. 
Therefore, to prevent the rate coefficient from growing too large at high $T$, 
the relation~(\ref{highrate})
is multiplied by an exponential damping factor to give Eq.~(\ref{highT}). 
The exact form is not important as the molecular abundances decrease rapidly 
for $T\gtrsim 5000$~K.

\section{Other details}

\subsection{Baseline models, unmodelled species}

Many species have level energies and transition probabilities tabulated in
NIST, but have no electron collisional rates at all.
For these species  we created ``baseline'' 
Stout data files. These contain the NIST level energies and transition probabilities  
but use the $\bar g$ approximation for all collision data.  
These are marked as ``baseline'' in Table \ref{TabDataSources}.

It was not possible to create models for all ions.
NIST did not have sufficient data to compute models 
for the following species: \ion{F}{1}, \ion{Cl}{13}, 
\ion{Cl}{15}, \ion{Sc}{1}, \ion{Ti}{1}, \ion{Ti}{2}, \ion{V}{1}, \ion{V}{2}, 
\ion{V}{3}, \ion{V}{5}, \ion{V}{14}, \ion{Cr}{1}, \ion{Cr}{3}, \ion{Cr}{5},
\ion{Mn}{2}, \ion{Mn}{3}, \ion{Mn}{4}, \ion{Mn}{7},
\ion{Co}{1}, \ion{Co}{4}, \ion{Co}{5}, \ion{Co}{6}, \ion{Co}{7}, \ion{Co}{9},
\ion{Ni}{6}, \ion{Ni}{8}, \ion{Ni}{10}, \ion{Cu}{2},  \ion{Cu}{3}, \ion{Cu}{4},
\ion{Cu}{5}, \ion{Cu}{6}, \ion{Cu}{7}, \ion{Cu}{8}, \ion{Cu}{9}, \ion{Cu}{10},
\ion{Cu}{12}, \ion{Cu}{12}, \ion{Cu}{19}, \ion{Cu}{20}, \ion{Zn}{1}, \ion{Zn}{3},
from  \ion{Zn}{5} to \ion{Zn}{19}, and  \ion{Zn}{21}, \ion{Zn}{22}, \ion{Zn}{26}.
As a result Cloudy calculations do not predict lines of these ions.
Calculating sufficient data for these species should be a high priority.

\subsection{Masterlists - specifying which database}

Cloudy uses a total of three databases, Stout, described here, along
with Chianti and LAMDA.
Each database has its own ``masterlist'', a file specifying the
species present in that version of the database.
The Stout masterlist file was used to derive Table \ref{TabDataSources}.

In a particular Cloudy calculation, each of these masterlist files will be read.
It is likely that a particular species is present in more than one database and
its masterlist file.
The  priorities for deciding which database to use are:
1) the H- and He-like isoelectronic sequences are always treated with our 
unified model,
2) Stout,
3) Chianti,
and 4) LAMDA.

\subsection{Suprathermal electrons}

When cosmic rays or Auger electrons enter neutral gas they create a population
of suprathermal electrons which ionize and excite the gas \citep{Spitzer1968}.
We solve for the population of these suprathermal electrons explicitly.

These electrons have an energy of typically 20 - 40 eV and can cause internal 
excitations of all atoms and molecules. We include this as a general excitation 
process using the Born approximation outlined by \citet{ShemanskyEtal85}.

\section{Discussion and Summary}

As described in the mandate for the development of Stout, it is now far easier to 
maintain and update the line database in Cloudy, and to add entirely new species.  
With the addition of the species given in Table \ref{TabDataSources} there are now far more
lines predicted than in previous versions of the code, producing far richer spectra.

\begin{figure}[!ht]
\includegraphics[width=\linewidth]{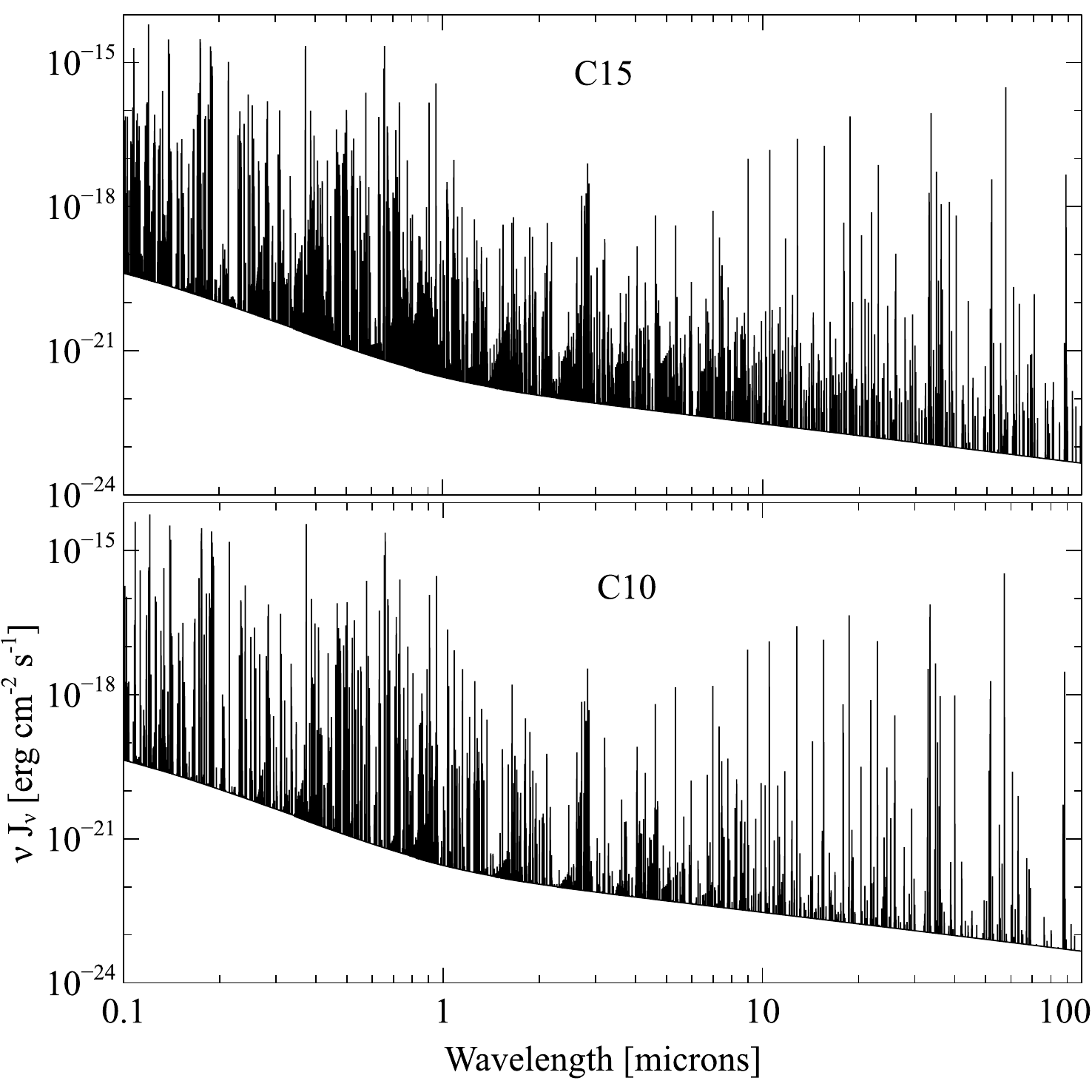}
\caption[Spectrum of $5\times \e{4}$~K metal-rich gas]
{The predicted spectrum of a $Z=35Z_{\sun}$ collisional gas
with a temperature of $5 \e{4} \K$. The upper panel shows 
the current results, for C15, while the lower panel is for 
C10, the last version before the move to external databases 
for ions. The density of lines is now far greater.
\label{fig:z30}}
\end{figure}

Figure \ref{fig:z30} shows an example.
This is a coronal equilibrium metal-rich ($Z = 35 Z_{\sun}$) gas with a density
of 1 cm$^{-3}$ and a temperature of $5\e{4} \K$.
The panels compare the current version, soon to be released as C15, 
with C10, the last Cloudy release before beginning the move of ionic models
to external databases.
There are now a far greater number of faint lines.

Despite the large increase in the number of lines the total cooling of the gas is
relatively unaffected.
\citet{Lykins.M12Radiative-cooling-in-collisionally-and-photo} describe
our calculation of the total gas cooling along with our strategy for 
determining how many levels to model.
The cooling is dominated by a few very strong lines so the large
number of faint lines do not increase it significantly.
The faint lines can be important when abundances are non-solar
(as in Figure \ref{fig:z30}), low-abundant species are of interest,
or if a number of faint lines blend to produce a stronger feature.

This paper is a definition of our database and explains how Cloudy uses it.
It is not intended as a review of the state of the art of atomic data in 2015.
Future papers will expand the atomic/molecular data using the framework
outlined here.

The database is designed to be easy to maintain and modify due to the need 
to constantly modify it as new data appear. The format follows the original 
data papers as closely as possible.

The methods we developed to fill in missing data are described. The data needs
of Cloudy are vast due to its very wide range of applicability.  We frequently
encounter cases where collisional rates are not available at all, or we need to
extrapolate beyond the range of computed data. The $\bar g$ approximation is 
used to provide missing electron collision data. This approximation has a very 
broad dispersion and we provide a method of checking on its impact on 
predictions. When rates or collision strengths are available, but we need to
extrapolate beyond the range of tabulated temperatures, we use physically 
motivated asymptotic limits. Tests show that predictions are mainly affected by 
the form of the low-$T$ extrapolation.

The Stout database is part of the Cloudy distribution, available on 
www.nublado.org. Its version number is the same as the Cloudy version number.
This paper is the defining documentation of Stout and should be cited if the 
database is used outside of Cloudy.

\acknowledgments
We thank the referee for a very helpful review of our manuscript.
GJF acknowledges support by NSF (1108928, 1109061, and 1412155), NASA 
(10-ATP10-0053, 10-ADAP10-0073, NNX12AH73G, and ATP13-0153), and STScI 
(HST-AR-13245, GO-12560, HST-GO-12309, GO-13310.002-A, and HST-AR-13914) 
and thanks to the Leverhulme Trust for support via the award of a Visiting 
Professorship at Queen�s University Belfast (VP1-2012-025).
RK research is funded by the European Social Fund under the Global Grant 
measure, project VP1-3.1-{\v S}MM-07-K-02-013. 

\appendix
\section{Data sources}

{\bf This Appendix describes the data sources currently used by the development version of Cloudy.
Species that are not explicitly listed in this Appendix use the Chianti database.
With the combination of these data, Chianti, and our special treatments of the H and He-like
iso-electronic sequences, Cloudy includes spectral models of all ions of the lightest
thirty elements.
}

%\LongTables
\renewcommand{\baselinestretch}{1.0}
\begin{deluxetable}{lll} 
\tabletypesize{\scriptsize}
\tablecolumns{3} 
\tablewidth{0pt}
\tabcolsep=1.5mm
%\tablenum{1}
\tablecaption{
Atomic data sources in Stout
\label{TabDataSources}
} 
\tablehead{ 
\colhead{$Z$} &
\colhead{Species} & 
\colhead{Data Source}\\
}
\startdata 
 3& Li\,{\sc i}   &baseline - see text     \\
 4& Be\,{\sc i}   &baseline      \\
  & Be\,{\sc ii}  &baseline      \\
 5& B\,{\sc i}    &baseline      \\ 
  & B\,{\sc ii}   &baseline      \\ 
  & B\,{\sc iii}  &baseline      \\ 
 6& C\,{\sc i}    &\citet{Johnson1987}, \citet{Mendoza1983},\\
  &               &\citet{Launay1977},\\
  &               &\citet{Roueff1990},\\
  &               &\citet{Schroder1991}, \citet{Staemmler1991}\\
  & C\,{\sc ii}   &\citet{Tayal2008}, \citet{Wiesenfeld2014}, \\
  &               &\citet{Goldsmith2012apjs}\\
  & C\,{\sc iii}  &\citet{Berrington1985}\\
 7& N\,{\sc i}    &\citet{FroeseFischer20041}, \citet{Tayal2000}\\
  & N\,{\sc v}    &\citet{Liang2011}\\
 8& O\,{\sc i}    &\citet{Bell1998}, \citet{Wang1992},\\
  &               & \citet{Barklem2007}, \citet{Abrahamsson2007},\\
  &               &\citet{Krems2006}, \citet{Monteiro1987},\\ 
  &               &\citet{Jaquet1992}, \citet{Pequignot1990} \\
  & O\,{\sc ii}   &\citet{Kisielius2009}, \citet{FroeseFischer20041}\\
 9& F\,{\sc i}    &baseline\\
  & F\,{\sc ii}   &\citet{Butler1994}\\
  & F\,{\sc iii}  &baseline\\
  & F\,{\sc iv}   &\citet{Lennon1994}\\
  & F\,{\sc v}    &baseline\\
  & F\,{\sc vi}   &baseline\\
  & F\,{\sc vii}  &baseline\\
10& Ne\,{\sc i}   &baseline\\
  & Ne\,{\sc ii}  &\citet{Griffin2001}, \citet{Hollenbach1989}\\
11& Na\,{\sc i}   &Verner private communication\\
  & Na\,{\sc ii}  &baseline\\
12& Mg\,{\sc i}   &\citet{Barklem2012}, \citet{Leep1976},\\
  &               & \citet{Mendoza1983}\\
  & Mg\,{\sc iii} &\citet{Liang2010}\\
13& Al\,{\sc i}   &baseline\\
  & Al\,{\sc iii} &\citet{Dufton1987}, \\
  &               &\citet{Sampson1990}\\
  & Al\,{\sc iv}  &baseline\\
  & Al\,{\sc vi}  &\citet{Butler1994}\\
14& Si\,{\sc i}   &\citet{Hollenbach1989}\\
  & Si\,{\sc ii}  &\citet{Tayal2008}, \citet{Dufton1994}, \\
  &               &\citet{Barinovs2005}\\
  & Si\,{\sc iii} &\citet{Dufton1989}, \\
  &               &\citet{Callegari1998}, \\
  &               &\citet{Dufton1983}\\
  & Si\,{\sc iv}  &\citet{Liang2009}\\
  & Si\,{\sc vii} &\citet{Butler1994}\\
  & Si\,{\sc ix}  &\citet{Lennon1994}\\
15& P\,{\sc i}    &{baseline}\\
  & P\,{\sc ii}   &\citet{Krueger1970}\\
  & P\,{\sc iii}  &\citet{Krueger1970}\\
  & P\,{\sc iv}   &baseline\\
  & P\,{\sc vi}   &baseline\\
16& S\,{\sc i}    &\citet{Hollenbach1989}\\
  & S\,{\sc ii}   &\citet{Kisielius2014}, \\
  &               &\citet{Tayal2010}\\
  & S\,{\sc iii}  &\citet{Hudson2012_SIII}\\
17& Cl\,{\sc i}   &\citet{Hollenbach1989}\\
  & Cl\,{\sc v}   &baseline\\
  & Cl\,{\sc vi}  &baseline\\
  & Cl\,{\sc vii} &\citet{Liang2009}\\
  & Cl\,{\sc viii}&\citet{Liang2010}\\
  & Cl\,{\sc ix}  &\citet{Berrington1998}\\
18& Ar\,{\sc i}   &{baseline}\\
  & Ar\,{\sc ii}  &\citet{Pelan1995}\\
  & Ar\,{\sc iii} &\citet{Galavis1995}\\
  & Ar\,{\sc iv}  &\citet{RamsbottomMNRAS1997}\\
  & Ar\,{\sc v}   &\citet{Galavis1995}\\
  & Ar\,{\sc vi}  &\citet{Saraph1996}\\
19& K\,{\sc i}    &{baseline}\\
  & K\,{\sc ii}   &{baseline}\\
  & K\,{\sc iii}  &\citet{Pelan1995}\\
  & K\,{\sc iv}   &\citet{Galavis1995}\\
  & K\,{\sc vii}  &\citet{Saraph1996}\\
  & K\,{\sc viii} &{baseline}\\
  & K\,{\sc x}    &{baseline}\\
20& Ca\,{\sc i}   &{baseline}\\
  & Ca\,{\sc iii} &{baseline}\\
  & Ca\,{\sc iv}  &\citet{Pelan1995}\\
  & Ca\,{\sc vi}  &{baseline}\\
21& Sc\,{\sc i}   &baseline\\
  & Sc\,{\sc ii}  &\citet{Wasson2011}\\
  & Sc\,{\sc iii} &baseline\\
  & Sc\,{\sc iv}  &baseline\\
  & Sc\,{\sc v}   &\citet{Pelan1995}\\
  & Sc\,{\sc vi}  &baseline\\
  & Sc\,{\sc vii} &baseline\\
  & Sc\,{\sc viii}&baseline\\
  & Sc\,{\sc ix}  &baseline\\
  & Sc\,{\sc ix}  &baseline\\
  & Sc\,{\sc x}   &baseline\\
  & Sc\,{\sc xi}  &baseline\\
  & Sc\,{\sc xii} &baseline\\
  & Sc\,{\sc xiii}&\citet{Saraph1994}\\
  & Sc\,{\sc xiv} &baseline\\
  & Sc\,{\sc xv}  &baseline\\
  & Sc\,{\sc xvi} &baseline\\
  & Sc\,{\sc xvii}&baseline\\
  & Sc\,{\sc xviii}&baseline\\
22& Ti\,{\sc iii} &baseline\\
  & Ti\,{\sc iv}  &baseline\\
  & Ti\,{\sc v}   &baseline\\
  & Ti\,{\sc vi}  &\citet{Pelan1995}\\
  & Ti\,{\sc vii} &baseline\\
  & Ti\,{\sc viii}&baseline\\
  & Ti\,{\sc ix}  &baseline\\
  & Ti\,{\sc x}   &baseline\\
  & Ti\,{\sc xiii}&baseline\\
23& V\,{\sc iv}   &baseline\\
  & V\,{\sc vi}   &baseline\\
  & V\,{\sc vii}  &\citet{Pelan1995}\\
  & V\,{\sc viii} &baseline\\
  & V\,{\sc ix}   &baseline\\
  & V\,{\sc x}    &baseline\\
  & V\,{\sc xi}   &baseline\\
  & V\,{\sc xii}  &baseline\\
  & V\,{\sc xiii} &baseline\\
%  & V\,{\sc xiv}  &baseline\\
  & V\,{\sc xv}   &\citet{Berrington1998}\\
  & V\,{\sc xvi}  &baseline\\
  & V\,{\sc xvii} &baseline\\
  & V\,{\sc xviii}&baseline\\
  & V\,{\sc xix}  &baseline\\
  & V\,{\sc xx}   &baseline\\
  & V\,{\sc xxi}  &baseline\\
24& Cr\,{\sc ii}  &\citet{Grieve2012}\\
  & Cr\,{\sc iii} &{baseline}\\
  & Cr\,{\sc iv}  &{baseline}\\
  & Cr\,{\sc v}   &{baseline}\\
%  & Cr\,{\sc vi}  &{baseline}\\
  & Cr\,{\sc x}   &{baseline}\\
  & Cr\,{\sc xi}  &{baseline}\\
  & Cr\,{\sc xii} &{baseline}\\
  & Cr\,{\sc xv}  &{baseline}\\
25& Mn\,{\sc i}   &{baseline}\\
  & Mn\,{\sc v}   &{baseline}\\
  & Mn\,{\sc vi}  &{baseline}\\
%  & Mn\,{\sc vii} &{baseline}\\
  & Mn\,{\sc xi}  &{baseline}\\
  & Mn\,{\sc xii} &{baseline}\\
  & Mn\,{\sc xiii}&{baseline}\\
  & Mn\,{\sc xiv} &{baseline}\\
  & Mn\,{\sc xvi} &{baseline}\\
26& Fe\,{\sc i}   &\citet{Hollenbach1989}\\
  & Fe\,{\sc ii}  &\citet{Verner1999}\\
  & Fe\,{\sc iii} &\citet{Zhang1996}, \citet{Kurucz2009AIPC.1171...43K}\\
  & Fe\,{\sc vii} &\citet{Witthoeft.M08Atomic-data-from-the-IRON} \\
27& Co\,{\sc ii}  &baseline\\
  & Co\,{\sc iii} &baseline\\
  & Co\,{\sc viii}&baseline\\
%  & Co\,{\sc ix}  &baseline\\
  & Co\,{\sc x}   &baseline\\
  & Co\,{\sc xi}  &\citet{Pelan1995}\\
  & Co\,{\sc xii} &baseline\\
  & Co\,{\sc xiii}&baseline\\
  & Co\,{\sc xiv} &baseline\\
  & Co\,{\sc xv}  &baseline\\
  & Co\,{\sc xvi} &baseline\\
  & Co\,{\sc xvii}&baseline\\
28& Ni\,{\sc i}   &\citet{Hollenbach1989}\\
  & Ni\,{\sc ii}  &\citet{Cassidy2011}\\
  & Ni\,{\sc iii} &baseline\\
  & Ni\,{\sc iv}  &baseline\\
  & Ni\,{\sc v}   &baseline\\
  & Ni\,{\sc vii} &baseline\\
  & Ni\,{\sc ix}  &baseline\\
  & Ni\,{\sc xvii}&\citet{Aggarwal2007},\\
  &               &\citet{Hudson2012_NiXVII}\\
29& Cu\,{\sc i}   &baseline\\
  & Cu\,{\sc xiii}&baseline\\
  & Cu\,{\sc xiv} &baseline\\
  & Cu\,{\sc xv}  &baseline\\
  & Cu\,{\sc xvi} &baseline\\
  & Cu\,{\sc xvii}&baseline\\
  & Cu\,{\sc xviii}&baseline\\
  & Cu\,{\sc xxi} &baseline\\
  & Cu\,{\sc xxii}&baseline\\
  & Cu\,{\sc xxiii}&baseline\\
  & Cu\,{\sc xxiv}&baseline\\
  & Cu\,{\sc xxv} &baseline\\
30& Zn\,{\sc ii}  &\citet{Kisielius2015}\\
  & Zn\,{\sc iv}  &baseline\\
\enddata
\end{deluxetable}
%\clearpage

\bibliography{bibliography2,LocalBibliography}

\end{document}